\documentclass[cameraready]{Interspeech}

\title{MOS-Bias: From Hidden Gender Bias to Gender-Aware Speech Quality Assessment}

\author[affiliation={1}, equalcontribution]{Wenze}{Ren}
\author[affiliation={1}, equalcontribution]{Yi-Cheng}{Lin}
\author[affiliation={2}]{Wen-Chin}{Huang}
\author[affiliation={3}]{Erica}{Cooper}
\author[affiliation={4}]{Ryandhimas}{E. Zezario}
\author[affiliation={4}]{Hsin-Min}{Wang}
\author[affiliation={1}]{Hung-yi}{Lee}
\author[affiliation={4}, correspondingauthor]{Yu}{Tsao}

\address{
    $^1$ National Taiwan University\\
    $^2$ Nagoya University\\
    $^3$ National Institute of Information and Communications Technology\\
    $^4$ Academia Sinica
}

\email{\{d14945014, f12942075, hungyilee\}@ntu.edu.tw, yu.tsao@citi.sinica.edu.tw}


\usepackage{comment}
\usepackage{makecell}
\usepackage{graphicx}
\usepackage{adjustbox}
\usepackage{multirow}
\usepackage{pifont}
\usepackage[table]{xcolor}

\begin{document}

\maketitle

\begin{abstract}
    The Mean Opinion Score (MOS) serves as the standard metric for speech quality assessment, yet biases in human annotations remain underexplored. We conduct the first systematic analysis of gender bias in MOS, revealing that male listeners consistently assign higher scores than female listeners—a gap that is most pronounced in low-quality speech and gradually diminishes as quality improves. This quality-dependent structure proves difficult to eliminate through simple calibration. We further demonstrate that automated MOS models trained on aggregated labels exhibit predictions skewed toward male standards of perception. To address this, we propose a gender-aware model that learns gender-specific scoring patterns through abstracting binary group embeddings, thereby improving overall and gender-specific prediction accuracy. This study establishes that gender bias in MOS constitutes a systematic, learnable pattern demanding attention in equitable speech evaluation.
\end{abstract}

\section{Introduction}
The Mean Opinion Score (MOS), derived from human listening tests, is the gold standard for subjective speech quality assessment in tasks such as Text-to-Speech (TTS)~\cite{hsu2025breezyvoice}, Voice Conversion (VC)~\cite{8706604}, and Speech Enhancement (SE). However, large-scale listening tests are costly and time-consuming, thus driving the development of automated MOS prediction models, such as MOSNet~\cite{lo2019mosnet}, SSL-MOS~\cite{cooper2022generalization}, UTMOS~\cite{saeki2022utmos,baba2024t05} and HighRateMOS~\cite{ren2025highratemos}, which can directly estimate speech quality from audio. These models have become indispensable speech quality assessment tools in modern speech generation tasks.


Despite significant progress in improving the accuracy of automated MOS predictions, the quality of the MOS labels themselves has received limited attention~\cite{leng2021mbnet, huang2022ldnet}. While biases in subjective listening tests are well-documented~\cite{zielinski2008some, rosenberg17_interspeech}, a critical yet overlooked source lies in the demographic composition of the listener pool. In particular, whether listener gender constitutes a systematic source of such bias remains largely unexplored. When male and female listeners hold systematically different perception standards~\cite{10.3389/fnhum.2023.1077409}, simple averaging produces a composite score that fails to accurately reflect either group—and may even inadvertently favor one gender's perception standards while neglecting the other's.
In this study, we systematically examine for the first time how listener gender influences Mean Opinion Score (MOS) ratings. Using the BVCC dataset, we reveal a consistent pattern: male listeners systematically assign higher scores, with this difference being most pronounced in low-quality speech and gradually diminishing as quality improves. However, standard MOS labels are derived by averaging all listener ratings without demographic calibration. This approach obscures meaningful differences in gender perception, and the resulting composite scores fail to accurately represent either group. Yet such averaged labels are widely regarded as gender-neutral benchmarks.
Our empirical research challenges this assumption. We found that automatic MOS prediction models trained solely on these average labels (without any gender information) consistently predicted scores closer to male listeners' ratings than female listeners', even though each audio sample in the training set had fewer male listeners than female listeners on average. This reveals that average MOS labels are not gender-neutral: they implicitly encode male-perceived standards, which models faithfully learn and propagate. To address this issue, we propose a gender-aware MOS prediction model that incorporates a parallel gender prediction branch. Rather than directly using listener gender labels as input, this model performs conditional processing based on abstracted binary group embeddings, enabling it to autonomously discover gender-specific perception patterns within the data. This approach enhances prediction accuracy both at the gender-specific and overall levels.


Our main contributions are summarized as follows:
\begin{enumerate}
    \item We provide the first systematic evidence that male 
    listeners consistently assign higher MOS scores than female 
    listeners, with the gap largest in low-quality speech.
    
    \item We show that averaged MOS labels and models trained 
    on them implicitly inherit a male-leaning perceptual bias.
    
    \item We propose a gender-aware model via an abstract binary 
    group embeddings, improving both overall and gender-specific 
    prediction accuracy.
    
\end{enumerate}

    
    
\section{Related Work}



\subsection{Bias in Human Annotation}
The impact of annotator demographics on annotation results has been extensively studied in Natural Language Processing (NLP) and Computer Vision (CV). In NLP, research has shown that annotators' gender, age, and cultural background introduce systematic biases in tasks such as sentiment analysis~\cite{al-kuwatly-etal-2020-identifying}, hate speech detection, and toxicity labeling~\cite{sap-etal-2022-annotators}. Similarly, studies in CV have demonstrated that annotators' demographic attributes affect the consistency of image annotations and the perception of offensive visual content~\cite{10.1145/3531146.3533216, liao2023transformer}. The broader literature on machine learning fairness further confirms that simple averaging of annotations tends to mask substantial discrepancies across demographic subgroups~\cite{davani-etal-2022-dealing}, leading to labels that overreact to the majority's views~\cite{mehrabi2021survey}.

Although demographic factors, such as listeners' expertise and linguistic background, have been considered in speech quality assessment in the Blizzard Challenge, these studies remain fragmented and lack a systematic approach. Notably, gender-related factors have been overlooked in speech assessment, despite strong evidence that gender influences auditory perception and evaluation behavior.


\subsection{Fairness in Speech Processing}
Previous research on fairness in speech processing has primarily focused on differences in model performance, such as the poor performance of automatic speech recognition systems for specific genders or dialect groups~\cite{harris-etal-2024-modeling, elghazaly2025exploring, fenu2020improving}, and the inconsistent quality produced by TTS systems under different speaker characteristics~\cite{williams2020comparison, puhach2025gets}. However, these studies only address the fairness of generation and recognition~\cite{borre2025explore, gorrostieta19_interspeech, wu2025evaluating, lin2024listen}, neglecting the fairness of evaluation. Whether the metrics used to evaluate these systems are fair, especially whether MOS labels can equally reflect the perceptions of all listener groups, remains a completely separate and neglected issue. This study explores this topic for the first time.
\section{Bias Analysis in MOS Annotations}
\subsection{Dataset and Toolkit}
We adopted the SHEET~\cite{huang2024mos} toolkit as our experimental framework, which supports MOS training and prediction across multiple public datasets. Among all datasets supported by SHEET, BVCC~\cite{cooper2021voices} is the only one providing both speaker and listener gender metadata, making it the only viable choice for our analysis. BVCC integrates publicly released samples from the Blizzard Challenge, Voice Conversion Challenge, and ESPnet-TTS, with each speech sample evaluated by eight listeners. The dataset is divided into 4,974 training speech samples, 1,066 development speech samples, and 1,066 test speech samples.

\subsection{Problem Formulation}
Let $r_i$ denote the rating assigned by listener $i$ to utterance $u$, and let $\mathcal{M}$ and $\mathcal{F}$ denote the sets of male and female listeners, respectively. The standard MOS is computed as:
\begin{equation}
    \text{MOS}(u) = \frac{1}{N}\sum_{i=1}^{N} r_i
\end{equation}
We define gender-specific scores as:
\begin{equation}
    \text{MOS}_M(u) = \frac{1}{|\mathcal{M}|}\sum_{i \in \mathcal{M}} r_i, 
    \quad 
    \text{MOS}_F(u) = \frac{1}{|\mathcal{F}|}\sum_{i \in \mathcal{F}} r_i
\end{equation}
The standard MOS is a weighted average of $\text{MOS}_M(u)$ and $\text{MOS}_F(u)$, with weights proportional to group size. If male and female listeners systematically differ in their ratings, this weighted average is closer to whichever group contributes more listeners. A single aggregated score, therefore, does not accurately represent the perceptual standards of either group. We quantify these differences in Sections 3.3 and 3.4, and propose a model that explicitly accounts for them in Section 5.

\subsection{Overall Gender Rating Difference}
We quantified gender-based scoring differences within the training set. As shown in Table \ref{tab:Listener_and_speaker_gender_mos}, male listeners consistently awarded higher scores than female listeners across all conditions: for male speakers, male listeners scored 2.925, while female listeners scored 2.822; for female speakers, the corresponding values were 3.065 and 2.964. This gap remained stable regardless of speaker gender. Notably, both listener groups rated female speakers higher, yet the pattern of “male listeners rating higher” persisted across all speaker gender conditions. This demonstrates that listener gender and speaker gender have independent influences on perceived quality.

To verify that this pattern was not due to sampling variation, we employed Welch's t-test, which is specifically designed to handle unbalanced sample sizes. As shown in Table \ref{tab:bvcc_training_welch_t_test}, the scoring differences between male and female listeners were statistically significant across all conditions, with all p-values falling well below conventional significance thresholds. This confirms that the observed gap reflects genuine structural differences rather than random variation.

It is worth noting that this finding cannot be attributed to gender imbalance within the listener group. In the training set, each audio clip received ratings from 3.6 male listeners and 4.25 female listeners. Despite this, the rating disparity persisted, further confirming its structural nature.

\begin{table}[t]
\fontsize{7}{9}\selectfont
\centering
\caption{The mean and standard deviation of MOS scores for listeners of different genders corresponding to speakers of different genders in the BVCC training set.}
\vspace{-8pt}
\label{tab:Listener_and_speaker_gender_mos}
\begin{tabular}{llccc}
\toprule
\textbf{Listener} & \textbf{Speaker} & \textbf{Mean} & \textbf{Std} & \textbf{Count} \\
\midrule
Male   & Male   & 2.925 & 1.137 & 9997  \\
Male   & Female & 3.065 & 1.205 & 8229  \\
Female & Male   & 2.822 & 1.175 & 11541 \\
Female & Female & 2.964 & 1.271 & 9598  \\
\bottomrule
\end{tabular}
\end{table}
\begin{table}[t]
\centering
\caption{Welch's $t$-test comparing MOS ratings between male and female listeners on the BVCC dataset. All differences are statistically significant ($p < 0.001$), indicating that male listeners consistently assign higher ratings than female listeners regardless of speaker gender.}
\label{tab:bvcc_training_welch_t_test}
\begin{tabular}{lcccc}
\toprule
\textbf{Condition} & \makecell{\textbf{Male}\\\textbf{Listener}} & \makecell{\textbf{Female}\\\textbf{Listener}} & $\boldsymbol{p}$ \\
\midrule
Male speaker   & 2.925 & 2.822 & $8.86 \times 10^{-11}$ \\
Female speaker & 3.065 & 2.964 & $5.35 \times 10^{-8}$  \\
Overall        & 2.988 & 2.886 & $4.58 \times 10^{-17}$ \\
\bottomrule
\end{tabular}
\end{table}
\subsection{Quality-Dependent Gender Difference}
\begin{figure}[t]
\centering
\caption{Across the four quality tiers of the BVCC training set, stratified by speaker gender, the mean rating difference between male and female listeners.}
\includegraphics[width=\columnwidth]{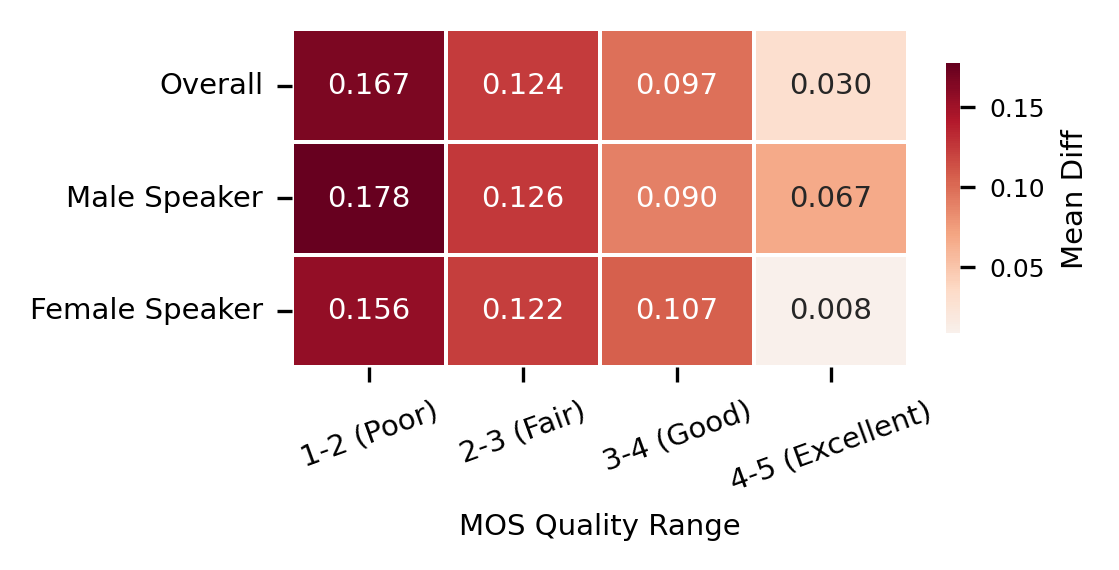}
\vspace{-6pt}
\vspace{-10pt}
\label{fig:BVCC_train_set_listener_diff_heatmap}
\end{figure}

To examine how the gender gap varies with speech quality, we divided the training set into four quality intervals based on overall MOS: 1–2 (Poor), 2–3 (Average), 3–4 (Good), and 4–5 (Excellent). For each interval, we computed the mean rating difference (male minus female) across all clips, stratified by speaker gender. Results are shown in Figure~\ref{fig:BVCC_train_set_listener_diff_heatmap}.

Averaged across speaker genders, the gap narrows monotonically with quality: 0.167 (1–2), 0.124 (2–3), 0.097 (3–4), and 0.030 (4–5). This pattern holds for each speaker gender, indicating that the quality-dependent gap is driven by listener gender and is not an artifact of speaker gender. The reduced gap at the highest quality level is consistent with a ceiling effect, though greater perceptual agreement on high-quality speech may also play a role.

This quality-dependent structure means that gender differences are not a fixed offset: a single global correction cannot adequately capture a gap whose magnitude varies systematically with speech quality. This motivates models that can learn gender-specific scoring patterns across quality conditions, which we explore in Sec.~\ref{sec:gender-aware-mos}.
\section{Limitations of Gender-Agnostic MOS}
We examine whether automated MOS prediction models, trained on these aggregated labels, produce predictions that differentially align with one gender group's ratings over the other.

\subsection{Experimental Setup}
We adopt SSL-MOS as the baseline model, as it represents the mainstream framework for automated MOS prediction and serves as the foundational architecture supporting follow-up models (including UTMOS) by combining additional features.

To ensure robustness, we trained the model under three independent random seeds (1337, 2337, 3337) with all other conditions held constant. The model was optimized using SGD ($\text{lr} = 1\times10^{-3}$) for up to 100,000 steps, with early stopping triggered after 20 consecutive evaluations without improvement in system-level SRCC.

\subsection{Bias Inheritance Analysis}

\begin{table}[t]
\fontsize{7}{9}\selectfont
\centering
\caption{\small Bias inheritance analysis on the testing set. Model predictions are evaluated against ground truth (GT) from all (All), male (M), and female (F) listeners.}
\label{tab:SSL_MOS_Baseline}
\footnotesize
\setlength{\tabcolsep}{3pt}
\begin{tabular}{l | cccc cccc}
\toprule
\multicolumn{1}{c|}{\multirow{2}{*}{\textbf{GT}}} & \multicolumn{4}{c}{\textbf{Utterance-Level}} & \multicolumn{4}{c}{\textbf{System-Level}} \\
\cmidrule(lr){2-5} \cmidrule(lr){6-9}
 & LCC & SRCC & MSE & KTAU & LCC & SRCC & MSE & KTAU \\
\midrule
All & \textbf{0.853} & \textbf{0.855} & \textbf{0.290} & \textbf{0.673} & \textbf{0.919} & \textbf{0.919} & \textbf{0.128} & \textbf{0.758} \\
M   & 0.806 & 0.811 & 0.372 & 0.628 & 0.901 & 0.899 & 0.141 & 0.727 \\
F   & 0.802 & 0.801 & 0.430 & 0.614 & 0.888 & 0.886 & 0.194 & 0.713 \\
\bottomrule
\end{tabular}
\end{table}


We evaluate predictions at both the utterance level and the system level. We report Linear Correlation Coefficient (LCC) and Spearman Rank Correlation Coefficient (SRCC) for correlation, Mean Squared Error (MSE) for prediction error, and Kendall's tau (KTAU) for ranking consistency. Table~\ref{tab:SSL_MOS_Baseline} presents results evaluated against three ground-truth references: overall MOS (All), male listener MOS (M), and female listener MOS (F), with all metrics averaged across three random seeds.

The model achieves its lowest error against the overall MOS it was trained on. When the same predictions are evaluated against gender-specific ground truths, we observe a consistent asymmetry: predictions are systematically closer to male listener ratings than to female listener ratings. Utterance-level MSE is 0.372 against male ground truth versus 0.430 against female ground truth (15.6\% relative gap); at the system level, MSE is 0.141 versus 0.194 (37.6\% relative gap). System-level LCC likewise favors male listeners (0.899 vs. 0.888).

Since the model receives no gender information during training, and its sole optimization target is the aggregated MOS, this asymmetry indicates that the aggregated label more closely approximates male perceptual standards than female ones. The prediction gaps in Table~\ref{tab:SSL_MOS_Baseline} empirically demonstrate a disparity that standard overall-only evaluation would not reveal. This motivates training objectives that explicitly account for gender-specific perception, which we explore in Sec.~\ref{sec:gender-aware-mos}.

\begin{table*}[t]
\centering
\fontsize{7}{9}\selectfont 
\newcommand{\std}[1]{{\tiny\color{gray}$\pm$#1}}
\caption{\small Performance comparison on the BVCC test set, grouped by Ground Truth listener set. All results are averaged over three independent seeds(1337, 2337, 3337).
\textbf{Prediction} denotes the branch output used for evaluation.
Bold indicates the best result within each group.
\colorbox{blue!10}{Shaded rows} highlight the proposed gender-specific branch predictions.}
\vspace{-6pt}
\label{tab:gender_mos_results}
\setlength{\tabcolsep}{4pt}
\begin{tabular}{l cc | cccc | cccc}
\toprule
\multirow{2}{*}{\textbf{Model}}
  & \multirow{2}{*}{\textbf{Prediction}}
  & \multirow{2}{*}{\textbf{Ground Truth}}
  & \multicolumn{4}{c|}{\textbf{Utterance-Level}}
  & \multicolumn{4}{c}{\textbf{System-Level}} \\
\cmidrule(lr){4-7}\cmidrule(lr){8-11}
 & & & LCC & SRCC & MSE & KTAU
     & LCC & SRCC & MSE & KTAU \\

\midrule
\multicolumn{11}{l}{\textit{Overall Prediction Quality (GT: All Listeners)}} \\
\midrule

Baseline
  & Avg & All
  & 0.853\std{0.002} & 0.855\std{0.001}
  & 0.290\std{0.004} & 0.673\std{0.002}
  & 0.919\std{0.004} & 0.919\std{0.005}
  & 0.128\std{0.001} & 0.758\std{0.006} \\

\rowcolor{blue!8}
Gender-MOS
  & Avg & All
  & \textbf{0.862}\std{0.003} & \textbf{0.862}\std{0.002}
  & \textbf{0.239}\std{0.016} & \textbf{0.683}\std{0.002}
  & \textbf{0.921}\std{0.005} & \textbf{0.918}\std{0.003}
  & \textbf{0.114}\std{0.020} & \textbf{0.761}\std{0.004} \\

\midrule
\multicolumn{11}{l}{\textit{Gender-Specific Prediction Quality (GT: Male Listeners)}} \\
\midrule

Baseline
  & Avg & Male
  & 0.806\std{0.002} & 0.811\std{0.001}
  & 0.372\std{0.004} & 0.628\std{0.002}
  & 0.901\std{0.001} & 0.899\std{0.003}
  & 0.141\std{0.001} & 0.727\std{0.003} \\

Gender-MOS
  & Avg & Male
  & 0.816\std{0.001} & 0.819\std{0.002}
  & 0.335\std{0.016} & 0.636\std{0.002}
  & 0.904\std{0.003} & 0.903\std{0.001}
  & 0.142\std{0.023} & \textbf{0.737}\std{0.002} \\

\rowcolor{blue!8}
Gender-MOS
  & Male & Male
  & \textbf{0.817}\std{0.001} & \textbf{0.819}\std{0.002}
  & \textbf{0.332}\std{0.012} & \textbf{0.636}\std{0.002}
  & \textbf{0.905}\std{0.003} & \textbf{0.903}\std{0.001}
  & \textbf{0.134}\std{0.017} & 0.736\std{0.002} \\

\midrule
\multicolumn{11}{l}{\textit{Gender-Specific Prediction Quality (GT: Female Listeners)}} \\
\midrule

Baseline
  & Avg & Female
  & 0.802\std{0.003} & 0.801\std{0.003}
  & 0.430\std{0.005} & 0.614\std{0.003}
  & 0.888\std{0.005} & 0.886\std{0.006}
  & 0.194\std{0.001} & 0.713\std{0.009} \\

Gender-MOS
  & Avg & Female
  & 0.806\std{0.004} & 0.806\std{0.000}
  & \textbf{0.366}\std{0.016} & 0.620\std{0.000}
  & 0.888\std{0.006} & 0.886\std{0.005}
  & \textbf{0.165}\std{0.018} & 0.711\std{0.005} \\

\rowcolor{blue!8}
Gender-MOS
  & Female & Female
  & \textbf{0.807}\std{0.005} & \textbf{0.807}\std{0.000}
  & \textbf{0.366}\std{0.028} & \textbf{0.621}\std{0.001}
  & \textbf{0.888}\std{0.007} & \textbf{0.887}\std{0.005}
  & 0.169\std{0.030} & \textbf{0.713}\std{0.006} \\

\bottomrule
\end{tabular}
\vspace{-10pt}
\end{table*}
\section{Gender-Aware MOS Prediction}
\label{sec:gender-aware-mos}
We have demonstrated that gender bias exists in both human annotations and models trained on them. We further pose a deeper question: Can modeling gender-specific MOS patterns improve prediction performance over a single-label model? To address this, we designed a gender-aware framework based on the SSL-MOS architecture and evaluated whether explicitly modeling gender-specific rating behaviors can improve prediction accuracy.

\subsection{Model Architecture}
\begin{figure}[t]
\centering
\caption{The proposed gender-aware MOS prediction architecture. A shared SSL encoder feeds into two networks: a Mean Net predicting overall MOS, and a Gender Net predicting gender-specific MOS scores with shared projection weights.}
\vspace{-6pt}
\includegraphics[width=\columnwidth]{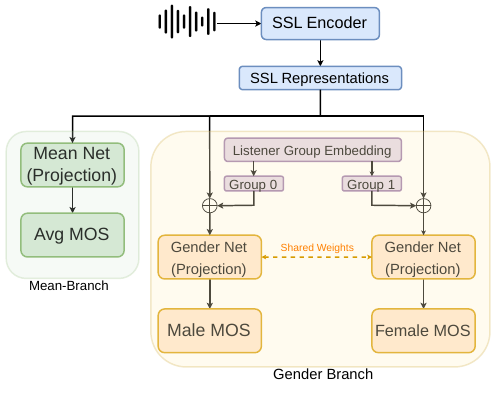}
\vspace{-32pt}
\label{fig:gender_mos_pipeline}
\end{figure}
We start with the baseline SSL-MOS model and introduce a parallel prediction branch to enable gender-differentiated MOS estimation. Specifically, as shown in Figure \ref{fig:gender_mos_pipeline}, we replicate the MOS prediction network to create a “gender-MOS” branch that operates in parallel with the original “mean-MOS” branch. This joint architecture is carefully designed: rather than training two completely independent models for male and female listeners (which would reduce the effective training data available to each), the shared SSL encoder enables both branches to benefit from the full dataset while still learning gender-specific rating patterns. Meanwhile, the key design decision is how to input listener gender information: we avoid directly concatenating gender labels (which would violate the baseline model's gender-neutral design) and instead condition the gender-MOS branch on two hard-coded binary values: 0 and 1. These values represent two abstract listener groups that the model must learn autonomously: Group 1 corresponds to male rating patterns, and Group 0 corresponds to female rating patterns. This design embodies a deliberate stance: the base model operates as a gender-agnostic listener, and we expect it to naturally develop gender perception through exposure to gendered real-world data signals (rather than explicit demographic labels). The final architecture produces three parallel outputs: the average MOS prediction for the original branch, the MOS prediction from the male listener's perspective for the gender MOS branch, and the MOS prediction from the female listener's perspective.

\subsection{Training Objectives}
We define the training objective as a multi-task loss composed of three weighted components:

\begin{equation*}
\mathcal{L}_{\text{total}} = \mathcal{L}_{\text{avg}} + \mathcal{L}_{\text{male}} + \mathcal{L}_{\text{female}}
\end{equation*}

where each term is the MSE between the corresponding branch prediction and its gender-specific ground truth. Equal weighting (1:1:1) 
ensures no single perspective dominates the optimization.

\subsection{Results and Analysis}
\noindent\textbf{Overall Performance.}
As shown in Table \ref{tab:gender_mos_results},When the ground truth is all listeners, the gender-aware model achieves better overall prediction quality than the baseline. At the utterance level, the gender-aware model achieves an LCC of 0.862 and an MSE of 0.239, while the baseline model achieves 0.853 and 0.290, respectively. This suggests that the auxiliary specificity objective may provide complementary training signals that assist the primary prediction branch rather than compete with it.

\noindent\textbf{Gender-Specific Accuracy.}
As shown in Table \ref{tab:gender_mos_results}, when the groundtruth is male or female listeners respectively, gender-specific branches improved prediction accuracy for each gender. For male listeners, the LCC for the utterance-set rose from 0.806 (baseline) to 0.817, and MSE dropped from 0.372 to 0.332. For female listeners, LCC increased from 0.802 to 0.807, and MSE fell from 0.430 to 0.366. The improvement for male listeners exceeded that for females, possibly reflecting stronger internal consistency in male ratings. Within the mean-branch of the gender MOS model, gender-specific prediction errors were also lower than in the baseline model (MSE decreased from 0.372 to 0.335 for males and from 0.430 to 0.366 for females). This shows that multi-task learning enhances both overall prediction accuracy and the modeling of gender pattern differences.

\noindent\textbf{Results Analysis.}
These results yield three key conclusions: First, the model successfully learned gender-specific scoring patterns solely through abstract binary encoding, the model successfully captures the systematic differences between the two predefined listener groups. Second, explicitly modeling gender-related variance enables the primary prediction branch to learn intrinsic speech quality more purely. Third, the availability of an independent gender prediction head lays the practical groundwork for future fairness-oriented interventions.
\section{Conclusion}
This study presents the first systematic investigation of gender bias in MOS-based speech quality assessment. We find that male listeners consistently assign higher ratings than female listeners, with the gap largest for low-quality speech (mean difference of 0.167) and diminishing to near zero at excellent quality, a quality-dependent structure that fixed calibration strategies cannot correct. We further show that this bias propagates into automated models: even without any gender supervision, model predictions align systematically closer to male ratings, with system-level MSE gaps reaching up to 37.6\% relative to female ground truth. To address this, we propose a gender-aware model conditioned on abstract binary group embeddings. Notably, the model successfully discovers gender-specific perceptual patterns without explicit demographic labels, improving both overall and gender-specific prediction accuracy. Future work will develop bias mitigation methods for MOS labels and models, and validate our approach across additional datasets. We expect this research to spur the learning community to critically examine fairness issues in speech quality assessment and advance toward more equitable evaluation practices.

\section{Generative AI Use Disclosure}

Generative AI tools assisted in the linguistic polishing of the manuscript. 
The authors remain solely responsible for the research design, experiments, analysis, and reported results. 
AI tools did not contribute to the substantive scientific content.


\bibliographystyle{IEEEtran}
\bibliography{mybib}

\end{document}